\documentclass[reprint,superscriptaddress,prl]{revtex4-1}
\usepackage{amsmath}
\usepackage{amssymb}
\usepackage{bm}
\usepackage{braket}
\usepackage{color}
\usepackage[varg]{txfonts}
\usepackage{varwidth}
\usepackage{dcolumn}
\usepackage[breaklinks,colorlinks=true,linkcolor=blue,urlcolor=cyan,citecolor=blue]{hyperref}
\usepackage{graphicx}

\begin{document}

\title{Rhombic skyrmion lattice coupled with orthorhombic structural distortion in EuAl$_{4}$}

\author{Masaki Gen}
\email{masaki.gen@riken.jp}
\affiliation{Department of Advanced Materials Science, University of Tokyo, Kashiwa 277-8561, Japan}
\affiliation{RIKEN Center for Emergent Matter Science (CEMS), Wako 351-0198, Japan}

\author{Rina Takagi}
\affiliation{RIKEN Center for Emergent Matter Science (CEMS), Wako 351-0198, Japan}
\affiliation{Institute of Engineering Innovation, University of Tokyo, Tokyo 113-0032, Japan}
\affiliation{Department of Applied Physics, University of Tokyo, Tokyo 113-8656, Japan}
\affiliation{PRESTO, Japan Science and Technology Agency (JST), Kawaguchi 332-0012, Japan}

\author{Yoshito Watanabe}
\affiliation{Department of Advanced Materials Science, University of Tokyo, Kashiwa 277-8561, Japan}

\author{Shunsuke Kitou}
\affiliation{RIKEN Center for Emergent Matter Science (CEMS), Wako 351-0198, Japan}

\author{Hajime Sagayama}
\affiliation{Institute of Materials Structure Science, High Energy Accelerator Research Organization, Tsukuba 305-0801, Japan}

\author{Naofumi Matsuyama}
\affiliation{Institute for Solid State Physics, University of Tokyo, Kashiwa 277-8581, Japan}

\author{Yoshimitsu Kohama}
\affiliation{Institute for Solid State Physics, University of Tokyo, Kashiwa 277-8581, Japan}

\author{Akihiko Ikeda}
\affiliation{Department of Engineering Science, University of Electro-Communications, Chofu, Tokyo 182-8585, Japan}

\author{Yoshichika \={O}nuki}
\affiliation{RIKEN Center for Emergent Matter Science (CEMS), Wako 351-0198, Japan}

\author{Takashi Kurumaji}
\affiliation{Department of Advanced Materials Science, University of Tokyo, Kashiwa 277-8561, Japan}

\author{Taka-hisa Arima}
\affiliation{Department of Advanced Materials Science, University of Tokyo, Kashiwa 277-8561, Japan}
\affiliation{RIKEN Center for Emergent Matter Science (CEMS), Wako 351-0198, Japan}

\author{Shinichiro Seki}
\affiliation{RIKEN Center for Emergent Matter Science (CEMS), Wako 351-0198, Japan}
\affiliation{Institute of Engineering Innovation, University of Tokyo, Tokyo 113-0032, Japan}
\affiliation{Department of Applied Physics, University of Tokyo, Tokyo 113-8656, Japan}

\begin{abstract}

The centrosymmetric tetragonal itinerant magnet EuAl$_{4}$ exhibits an intricate magnetic phase diagram including rhombic and square skyrmion-lattice (SkL) phases in the external magnetic field.
Here, we report a multi-axis dilatometric investigation of EuAl$_{4}$ by means of a newly designed fiber-Bragg-grating technique complemented by a resonant x-ray scattering experiment, revealing anisotropic magnetostriction and magnetovolume effect associated with successive phase transitions.
The rhombic and square SkL phases are found to possess $\sim$0.10\% and $\sim$0.03\% orthorhombic structural distortion within the $ab$ plane, respectively.
We propose that the coupling between the spin system and the lattice deformation should be essential for the structural instability in EuAl$_{4}$, yielding a rich variety of topological spin textures with spontaneous rotational-symmetry breaking as well as a potential controllability of the SkL phases by uniaxial stress or pressure.

\end{abstract}

\date{\today}
\maketitle

{\it Introduction.---} Spontaneous rotational-symmetry breaking (SRSB) has long been a central issue in condensed-matter physics.
A prominent recent topic is the electronic nematicity observed in heavy-fermion compounds \cite {2011_Oka}, topological kagome metals \cite{2022_Nie}, and copper/iron-based high-$T_{\rm c}$ superconductors \cite{2010_Law, 2012_Kas}, where the contribution of electron-phonon coupling has often been controversial.
The SRSB is more widely seen with the onset of a magnetic long-range order.
In the presence of magnetic frustration, the mutual coupling of spin and lattice degrees of freedom can induce a magnetostructural transition with crystal symmetry lowering \cite{2011_Lac}, as observed in solid oxygen \cite{1981_Fot, 1986_Ste} and pyrochlore-based antiferromagnets \cite{2000_Lee, 2007_Mat}.

A magnetic skyrmion, a particlelike swirling spin texture, offers a fertile playground to explore emergent electromagnetic responses and transport properties \cite{2009_Neu, 2010_Jon, 2012_Sch, 2017_Hsu, 2020_Hir, 2021_Aka}.
In bulk crystals, skyrmions are usually arranged periodically, forming a skyrmion lattice (SkL) with high rotational symmetry.
In the framework of the Ginzburg--Landau theory, a triangular SkL, characterized by a triple-${\mathbf Q}$ modulation with hexagonal symmetry, can be stabilized by the external magnetic field and entropy effect \cite{1989_Bog, 1994_Bog, 2009_Muh}.
This was observed in a number of chiral \cite{2009_Muh, 2010_Yu, 2011_Yu, 2012_Sek, 2015_Tok, 2019_Kan} and polar \cite{2015_Kez, 2017_Kur} magnets with the Dzyaloshinskii--Moriya (DM) interaction.
Subsequently, the SkL was also discovered in rare-earth-based centrosymmetric itinerant magnets \cite{2019_Kur, 2019_Hir, 2020_Kha, 2022_Kha, 2022_Tak}, whose magnetism is governed by the Ruderman-Kittel-Kasuya-Yosida (RKKY) interaction.
Several mechanisms such as thermal fluctuations \cite{2012_Oku}, higher-order multiple-spins interaction \cite{2017_Oza, 2017_Hay}, single-ion anisotropy \cite{2015_Leo, 2019_Hay, 2020_Wan}, dipole--dipole interaction \cite{2021_Ute, 2022_Pad}, double-exchange mechanism \cite{2021_Kat}, and interorbital frustration \cite{2020_Nom} are proposed as key ingredients for stabilizing the SkL in these compounds.
It should be noted that the RKKY interaction is highly dependent on the shape of the Fermi surface.
In fact, skyrmions in a tetragonal magnet GdRu$_{2}$Si$_{2}$ are arranged not in a triangular, but a square lattice \cite{2020_Kha, 2022_Kha}.

\begin{figure}[t]
\centering
\includegraphics[width=\linewidth]{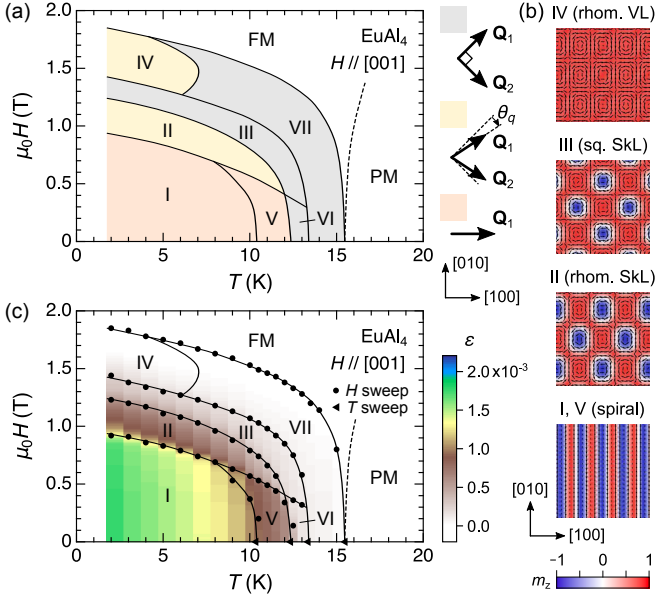}
\caption{(a) $H$-$T$ phase diagram of EuAl$_{4}$ for $H \parallel [001]$, which is an excerpt from Refs.~\cite{2022_Tak, 2022_Mei} and reproduced in the present work. PM and FM represent paramagnetic and ferromagnetic phases, respectively. The fundamental modulation vectors in each phase are depicted on the right side. (b) Schematics of the real-space spin configurations in phases I$\sim$V. (c) Contour plots of the orthorhombic structural distortion $\varepsilon = (b-a)/a_{0}$ mapped on the $H$-$T$ phase diagram revealed in this study. $\varepsilon$ is set to zero at 0~T and 18~K. Closed circles (triangles) indicate transition fields (temperatures at 0~T) determined from peaks in $\partial \varepsilon/\partial H$ ($\partial \varepsilon/\partial T$). The phase boundary between phases IV and VII was not resolved in our magnetostriction measurements.}
\label{Fig1}
\end{figure}

Recently, the SRSB of SkL was discovered in a binary intermetallic EuAl$_{4}$ \cite{2022_Tak}, which crystallizes in a centrosymmetric tetragonal structure (space group $I4/mmm$) with a square lattice of localized spin-7/2 Eu$^{2+}$ ions.
Figure~\ref{Fig1}(a) shows the magnetic-field-versus-temperature ($H$-$T$) phase diagram for $H \parallel [001]$, where seven magnetic phases (I--VII) appear, as well as the paramagnetic (PM) and the forced ferromagnetic (FM) phases \cite{2022_Tak, 2022_Mei}.
In zero magnetic field, four magnetic transitions from the PM phase to phases VII, VI, V, and I take place at $T_{\rm N1} = 15.4$~K, $T_{\rm N2} = 13.2$~K, $T_{\rm N3} = 12.2$~K, and $T_{\rm N4} = 10.0$~K, respectively \cite{2022_Tak, 2022_Mei, 2015_Nak, 2019_Shi, 2021_Sha, 2021_Kan, 2022_Zhu}.
The former two are of the second order, whereas the latter two are of the first order.
A single-crystal x-ray diffraction (XRD) study revealed a tetragonal-to-orthorhombic structural transition with the $B_{1g}$-type distortion at $T_{\rm N3}$ \cite{2019_Shi}.
When applying a magnetic field along [001], successive metamagnetic transitions from phase I to II, III, and IV take place  below 5~K.
A small-angle neutron scattering experiment \cite{2022_Tak} unveiled the magnetic structures of all these phases: phases I and V are single-${\mathbf Q}$ screw spiral with ${\mathbf Q}_{1}=(0.19, 0, 0)$ and (0.17, 0, 0), respectively; phases II and III are double-${\mathbf Q}$ SkL; phases IV and VII are double-${\mathbf Q}$ vortex-antivortex lattice (VL); and phase VI is double-${\mathbf Q}$ meron-antimeron lattice (ML).
Two fundamental modulation vectors in phases III, VI, and VII are ${\mathbf Q}_{1}=(q, q, 0)$ and ${\mathbf Q}_{2}=(q, -q, 0)$, where $q \sim 0.085$, so that the spin textures possess four-fold symmetry.
In phases II and IV, on the other hand, ${\mathbf Q}_{1}$ and ${\mathbf Q}_{2}$ are tilted by $\theta_{q} \sim 5^{\circ}$ toward the [100] direction, resulting in the rhombic SkL and VL, respectively [Fig.~\ref{Fig1}(b)].
In short, the SRSB is observed even in the double-${\mathbf Q}$ states (II and IV) as well as in the single-${\mathbf Q}$ states (I and V).

Theoretically, a {\it rectangular} SkL with four-fold symmetry breaking can appear on a square-lattice spin model by incorporating the competition among exchange interactions in momentum space $J_{\mathbf Q}$ with ${\mathbf Q}_{1} = (q, q, 0)$, ${\mathbf Q}_{2} = (q, -q, 0)$, and their higher harmonics, ${\mathbf Q}_{1} + {\mathbf Q}_{2}$ and ${\mathbf Q}_{1} - {\mathbf Q}_{2}$ \cite{2022_Hay_JPSJ, 2022_Hay}, though this picture is incompatible with the {\it rhombic} SkL with tilted two ${\mathbf Q}$ vectors in EuAl$_{4}$.
Recalling the structural transition at $T_{\rm N3}$, it is natural to anticipate that the spin-lattice coupling is responsible for stabilizing the rhombic double-${\mathbf Q}$ modulation in phases II and IV.
To shed light on this, it is important to clarify the in-field crystal-structure changes.
Dilatometry using the capacitance method has sufficient sensitivity, and a previous study \cite{2022_Mei} succeeded in observing the relative change in the lattice constant $a$ on a detwinned crystal of EuAl$_{4}$ while the information on the lattice constants $b$ and $c$ was lacking (we define $a < b$ below $T_{\rm N3}$).
Besides, the correspondence between the orthorhombic distortion and ${\mathbf Q}$ vectors has yet been revealed.

In this Letter, we investigate the lattice constant changes for three principal axes associated with the field-induced phase transitions in EuAl$_{4}$, unveiling anisotropic magnetostriction and magnetovolume effect.
Figure~\ref{Fig1}(c) shows a contour plot of the orthorhombic structural distortion $\varepsilon$ mapped on the $H$-$T$ phase diagram: e.g., $\sim$0.10\% and $\sim$0.03\% structural distortion is found at 2~K in the rhombic and square SkL phases, respectively.
We also performed a resonant x-ray scattering (RXS) experiment, revealing that the ${\mathbf Q}$ vector is oriented along the elongated $b$ axis in the single-${\mathbf Q}$ spiral phase (phase~I).
Based on the microscopic consideration of a spin model, we propose that the spin-lattice coupling should act as the principal driving force for the multiple magnetostructural transitions and the SRSB of SkL in EuAl$_{4}$.

\begin{figure}[t]
\centering
\includegraphics[width=\linewidth]{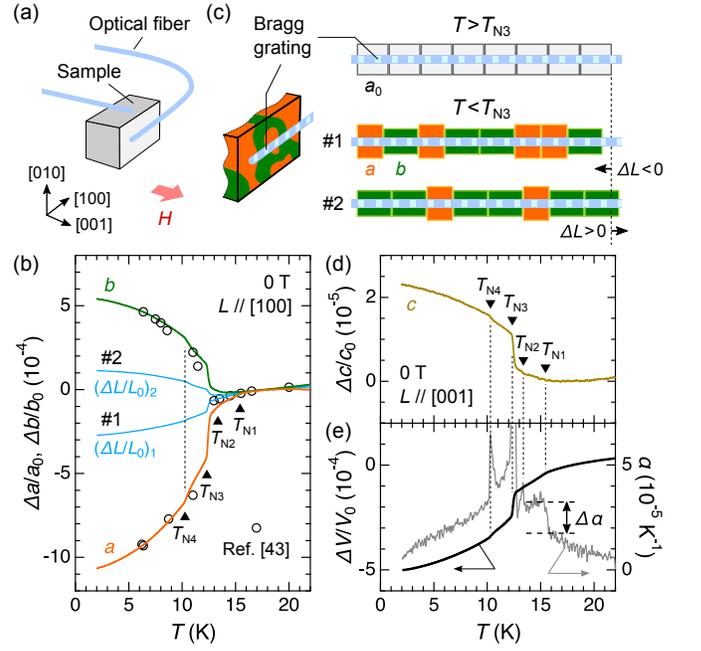}
\caption{(a)~Schematic of the sample setting of FBG experiments, where $\Delta L/L_{0}$ along [100] and [001] are simultaneously measured by adhering two optical fibers to one crystal orthogonally. (b)~Thermal expansion measured along [100] in zero field (cyan lines): $(\Delta L/L_{0})_{1}$ [$(\Delta L/L_{0})_{2}$] for setting~1 (setting~2). Open circles are extracted from the lattice constants $a$ and $b$ reported in Ref.~\cite{2019_Shi}. $\Delta a/a_{0}$ and $\Delta b/b_{0}$ are calculated by $(\Delta L/L_{0})_{1}$, $(\Delta L/L_{0})_{2}$, and the volume ratio of the crystallographic domains $(1-p_{i}) : p_{i}$ with $p_{1} = 0.49$ ($p_{2} = 0.74$) for setting~1 (setting~2). (c)~Schematic of the crystallographic domains distribution in the (001) plane below $T_{\rm N3}$. (d)~Thermal expansion measured along [001] in zero field ($\Delta c/c_{0}$). (e)~Volume thermal expansion (left axis) and its temperature derivative (right axis) in zero field. All the thermal expansion data were obtained in the warming process.}
\label{Fig2}
\end{figure}

{\it Methods.---} Single crystals of EuAl$_{4}$ were synthesized by the Al self-flux method as in Ref.~\cite{2022_Tak}.
The as-grown crystals were cut into the rectangular parallelepiped shape.
The magnetization was measured using a commercial magnetometer (Magnetic Property Measurement System, Quantum Design).
The dilatometry measurements were performed by the fiber-Bragg-grating (FBG) method using an optical sensing instrument (Hyperion si155, LUNA).
The relative sample-length changes $\Delta L/L_{0}$ along [100] and [001] were simultaneously measured, as illustrated in Fig.~\ref{Fig2}(a).
A sample with two FBGs was loaded in a cryostat equipped with a superconducting magnet (Spectromag, Oxford Instruments).
For experimental details, see the Supplemental Material \cite{SM}.
Throughout the Letter, we define $\Delta k$ ($k = L, a, b, c$) as $\Delta k \equiv k - k_{0}$ with the baseline value $k_{0}$ at 0~T and 18~K and assume $a_{0} = b_{0}$.
The RXS measurement was performed at BL-3A, Photon Factory, KEK, Japan, by using incident x-rays in resonance with an Eu $L_{2}$ absorption edge (7.615~keV).
A crystal with a flat (100) plane was attached on an Al plate using GE varnish and loaded into a cryostat equipped with a vertical-field superconducting magnet, where the scattering plane was set to be $(H, K, 0)$.
In all the experiments, a magnetic field was applied along [001].

{\it Results.---} The zero-field thermal expansion profiles measured along [100], $(\Delta L/L_{0})_{1}$ and $(\Delta L/L_{0})_{2}$, which were taken in independent sample settings, are shown by cyan lines in Fig.~\ref{Fig2}(b), along with the temperature evolution of $\Delta a/a_{0}$ and $\Delta b/b_{0}$ revealed by the previous XRD study \cite{2019_Shi} (open circles).
Opposite behaviors are observed between $(\Delta L/L_{0})_{1}$ and $(\Delta L/L_{0})_{2}$ below $T_{\rm N3}$; a positive thermal expansion for $(\Delta L/L_{0})_{1}$, while negative for $(\Delta L/L_{0})_{2}$.
This discrepancy should be attributed to the difference in the crystallographic-domain patterns around the local area where the FBG was glued in each setting [Fig.~\ref{Fig2}(c)].
Assuming $(\Delta L/L_{0})_{i}=(1-p_{i})(\Delta a/a_{0})+p_{i}(\Delta b/b_{0})$, we determine $p_{1} = 0.49$ for $(\Delta L/L_{0})_{1}$ and $p_{2} = 0.74$ for $(\Delta L/L_{0})_{2}$ so that they match the XRD data at 6.3~K.
Importantly, $p_{i}$ is reproduced no matter how many times the temperature or magnetic field is repeatedly scanned in our experiments (Fig.~S2 \cite{SM} in the Supplemental Material).
Thanks to this feature, we can decompose $\Delta a/a_{0}$ and $\Delta b/b_{0}$ from the two experimental data sets $(\Delta L/L_{0})_{1}$ and $(\Delta L/L_{0})_{2}$ in the whole measured $H$-$T$ region using the following relations: $\Delta a/a_{0} = [p_{2}(\Delta L/L_{0})_{1}-p_{1}(\Delta L/L_{0})_{2}]/(p_{2}-p_{1})$ and $\Delta b/b_{0} = [-(1-p_{2})(\Delta L/L_{0})_{1}+(1-p_{1})(\Delta L/L_{0})_{2}]/(p_{2}-p_{1})$.
The calculated $\Delta a/a_{0}$ and $\Delta b/b_{0}$ as a function of temperature are shown by orange and green lines in Fig.~\ref{Fig2}(b), respectively, which agree well with the XRD data.

As shown in Fig.~\ref{Fig2}(d), the observed $\Delta L/L_{0}$ along [001], which corresponds to $\Delta c/c_{0}$, exhibits negative thermal expansion below $T_{\rm N1}$.
The entire change in $\Delta c/c_{0}$ across the phase transitions is much smaller than those in $\Delta a/a_{0}$ and $\Delta b/b_{0}$.
This trend is in contrast to GdRu$_{2}$Si$_{2}$ \cite{2006_Pro} and suggests the weak out-of-plane spin-lattice coupling in EuAl$_{4}$.
We double check the reliability of our measurements by estimating thermodynamic quantities.
Figure~\ref{Fig2}(e) shows the volume thermal expansion calculated as $\Delta V/V_{0} = \Delta a/a_{0} + \Delta b/b_{0} + \Delta c/c_{0}$ and its temperature derivative $\alpha \equiv \partial (\Delta V/V_{0})/\partial T$.
$\alpha$ jumps by $\Delta \alpha \approx -1.5 \times 10^{-4}$~K$^{-1}$ at $T_{\rm N1}$.
By adopting this value in combination with the reported specific heat change $\Delta C_{p} \approx -6$~J/(K$\cdot$mol) at $T_{\rm N1}$ \cite{2015_Nak, 2022_Mei} and the volume $V \approx 107.1$~$\AA^{3}$/f.u. at 20~K \cite{2022_Ram} to the Ehrenfest relation ${\partial T_{\rm N}}/{\partial p} = T_{\rm N}V({\Delta \alpha}/{\Delta C_{p}})$, the pressure $p$ dependence of $T_{\rm N1}$ is estimated to 2.5~K/GPa.
This estimation agrees well with the previously obtained value ${\partial T_{\rm N1}}/{\partial p} = 2.24$~K/GPa \cite{2015_Nak}.

\begin{figure}[t]
\centering
\includegraphics[width=\linewidth]{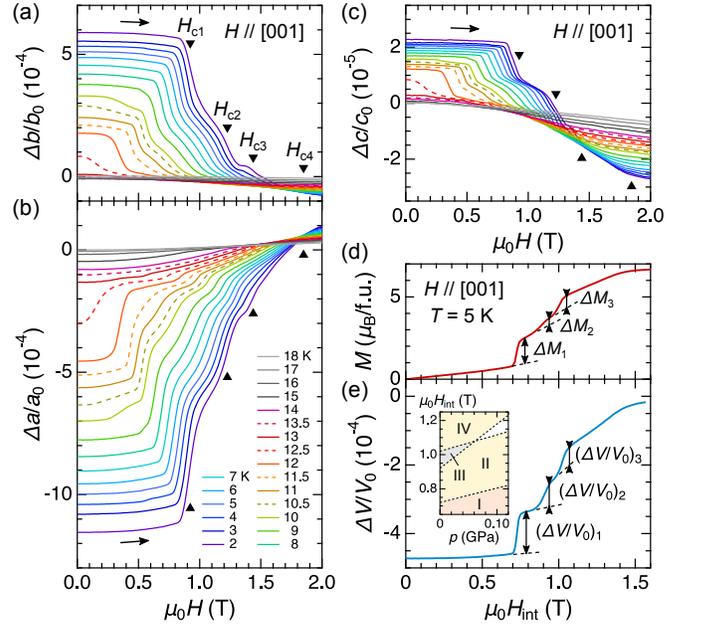}
\caption{[(a)--(c)] Magnetic-field evolution of the lattice constant changes (a) $\Delta b/b_{0}$, (b) $\Delta a/a_{0}$, and (c) $\Delta c/c_{0}$ for $H \parallel [001]$. The color scale of temperature is shown in the inset of (b). Triangles denote magnetic transitions at 2~K. [(d) and (e)] (d) Magnetization and (e) volume magnetostriction curves for $H \parallel [001]$ at 5~K. The horizontal axis $H_{\rm int}$ represents the magnetic field after demagnetization correction. The inset in (e) shows the thermodynamically predicted $H$-$p$ phase diagram. All the magnetostriction and magnetization data were obtained in the field-increasing process.}
\label{Fig3}
\end{figure}

Having confirmed the validity of our experimental and analytical methods from the thermal expansion data, we investigate the field-induced crystal-structure changes of EuAl$_{4}$.
Figures~\ref{Fig3}(a)--\ref{Fig3}(c) show magnetostriction curves for $H \parallel [001]$ measured at various temperatures.
We obtain the field evolution of $\Delta a/a_{0}$ and $\Delta b/b_{0}$ by the same procedure described above (Fig.~S1 \cite{SM} in the Supplemental Material).
As the magnetic field increases at 2~K, $\Delta a/a_{0}$ monotonically increases accompanied by jumps at first-order transitions at $H_{\rm c1}$, $H_{\rm c2}$, and $H_{\rm c3}$ in accord with Ref.~\cite{2022_Mei}.
In contrast, $\Delta b/b_{0}$ and $\Delta c/c_{0}$ monotonically decrease, indicating that the structural distortion associated with the magnetic ordering is gradually suppressed toward higher fields.
Furthermore, we performed similar sets of thermal expansion and magnetostriction measurements for $\Delta L/L_{0} \parallel [100]$ on a crystal nearly detwinned by applying the thermal stress (Fig.~S3 \cite{SM} in the Supplemental Material); the observed $\Delta L/L_{0}$ is close to $\Delta b/b_{0}$ and reasonably consistent with Figs.~\ref{Fig2}(b) and \ref{Fig3}(a), while a uniaxial stress effect is seen in a shift of structural transition temperatures $T_{\rm N3}$ and $T_{\rm N4}$ by more than 1~K.
$\varepsilon$ calculated by $\Delta b/b_{0} - \Delta a/a_{0}$ is visualized on the $H$-$T$ phase diagram in Fig.~\ref{Fig1}(c).
Note that $\varepsilon$ may be slightly underestimated in high-field and high-temperature sides due to the missing incorporation of possible slight orthorhombic distortion at 0~T and 18~K \cite{2022_Ram} in our analysis (for details, see the Supplemental Material \cite{SM}).

\begin{figure}[t]
\centering
\includegraphics[width=\linewidth]{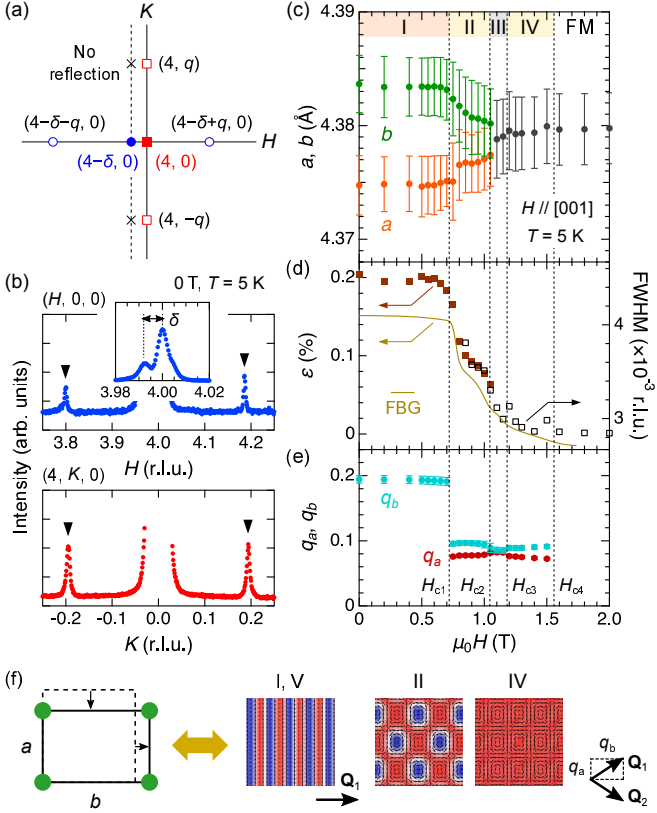}
\caption{Results of the RXS experiment performed at 5~K. (a) Schematic of the observed fundamental and magnetic Bragg reflections in zero field drawn on the scattering plane. (b) Intensity profiles in the $(H, 0, 0)$ and $(4, K, 0)$ scans in zero field. The inset shows the splitting pattern of the fundamental 400 reflection. [(c)--(e)] Magnetic-field evolutions of (c) the lattice constants $a$ and $b$, (d) the orthorhombic structural distortion $\varepsilon$ obtained from the RXS (symbols) and FBG experiments (solid line) \cite{comment1}, and (e) the $a$ and $b$ components of the ${\mathbf Q}$ vector. In (c) and (e), vertical bars represent ranges on real space corresponding to FWHMs of the fitting functions for the RXS intensity profiles \cite{comment2}. In (d), filled squares indicate $\varepsilon$ derived from $a$ and $b$ shown in (c) (left axis), and open squares indicate the FWHM of a single Lorentzian function fitted on the 400 peak (right axis). (f) Correspondence between the orthorhombic distortion and magnetic structure in phases I, II, IV, and V.}
\label{Fig4}
\end{figure}

Our magnetostriction data provide insights on the pressure effect on the stability of the rhombic and square SkL phases.
Figures~\ref{Fig3}(d) and \ref{Fig3}(e) show magnetization and volume magnetostriction curves, respectively, for $H \parallel [001]$ at 5~K.
Here, the demagnetization is corrected so as to accurately estimate the magnitudes of the magnetization jumps at $H_{\rm c1}$, $H_{\rm c2}$, and $H_{\rm c3}$: $\Delta M_{1} \approx 1.6$~$\mu_{\rm B}$/f.u., $\Delta M_{2} \approx 0.3$~$\mu_{\rm B}$/f.u., and $\Delta M_{3} \approx 0.6$~$\mu_{\rm B}$/f.u., respectively.
The corresponding volume jumps are $(\Delta V/V_{0})_{1} \approx 1.2 \times 10^{-4}$, $(\Delta V/V_{0})_{2} \approx 0.7 \times 10^{-4}$, and $(\Delta V/V_{0})_{3} \approx 0.5 \times 10^{-4}$, respectively.
According to Clausius--Clapeyron's equation ${\partial (\mu_{0}H_{\rm c})}/{\partial p} = {\Delta V}/{\Delta M}$, the pressure dependence of each critical field is estimated to ${\partial (\mu_{0}H_{\rm c1})}/{\partial p} \approx 0.8$, ${\partial (\mu_{0}H_{\rm c2})}/{\partial p} \approx 2.6$, and ${\partial (\mu_{0}H_{\rm c3})}/{\partial p} \approx 1.0$~T/GPa.
A relatively large value of ${\partial (\mu_{0}H_{\rm c2})}/{\partial p}$ is attributed to the substantial volume expansion at the transition to phase III in spite of the small magnetization jump.
The predicted $H$-$p$ phase diagram is shown in the inset of Fig.~\ref{Fig3}(e), suggesting that the square SkL can be annihilated by applying hydrostatic pressure lower than 0.1~GPa.
The stability of each SkL phase may also be controllable by tuning the chemical pressure such as the isovalent Ga substitution for Al \cite{2018_Sta}, which should act as the negative pressure.
Indeed, the $H$-$T$ phase diagrams of EuGa$_{4}$ and EuGa$_{2}$Al$_{2}$ are simpler than that of EuAl$_{4}$ \cite{2022_Zha, 2022_Moy}, presumably suggesting the absence of orthorhombic structural distortion and the rhombic SkL phase in these compounds.
A systematic investigation on the phase diagram of Eu(Ga$_{1-x}$Al$_{x}$)$_{4}$ system would be intriguing.

To reveal the one-to-one correspondence between the orthorhombic structural distortion and magnetic modulation, we performed the RXS experiment to observe both the fundamental 400 and magnetic Bragg reflections. 
Figure~\ref{Fig4}(a) shows the schematic of the observed reflections around (4, 0, 0) on the $(H, K, 0)$ scattering plane at 5~K in zero field.
The corresponding intensity profiles in the $(H, 0, 0)$ and $(4, K, 0)$ scans are shown in the upper and lower panels, respectively, in Fig.~\ref{Fig4}(b).
As shown in the inset of Fig.~\ref{Fig4}(b), we observe a peak splitting for the 400 reflection.
The estimated lattice constants are $a=4.3748$~$\AA$ and $b=4.3836$~$\AA$, yielding $\varepsilon = 2.0 \times 10^{-3}$.
This value is a bit larger than that obtained in the previous XRD, $\varepsilon = 1.6 \times 10^{-3}$ \cite{2019_Shi}, presumably owing to the extrinsic strain caused by the thermal expansion mismatch between the EuAl$_{4}$ crystal and the Al substrate in our RXS experiment.
Importantly, the magnetic peaks are observed at $(4-\delta-q, 0, 0)$, $(4-\delta+q, 0, 0)$, $(4, -q, 0)$, and $(4, q, 0)$ with $\delta = 0.002$ and $q = 0.194$.
This indicates that the magnetic peaks at the former (latter) two originate from the fundamental peak observed at $(4-\delta, 0, 0)$ [$(4, 0, 0)$].
Accordingly, we conclude that the ${\mathbf Q}$ vector in phase~I is oriented along the elongated $b$ axis.

Figures~\ref{Fig4}(c)--(e) summarize the field evolutions of the crystal structure in the $ab$ plane and the corresponding magnetic modulations, both of which are obtained during the same field-increasing process in the RXS experiment.
In Fig.~\ref{Fig4}(e), $q_{a}$ ($q_{b}$) represent the $a$ ($b$) component of ${\mathbf Q}_{1}$, as shown in Fig.~\ref{Fig4}(f) ($q_{a} = 0$ in phase~I).
The observed $q_{a}$ and $q_{b}$ well reproduce the previous neutron study \cite{2022_Tak}.
Below $\mu_{0}H_{\rm c2} \approx 1.05$~T, the magnitude of the 400 peak splitting is larger than each peak width so that we can derive $\varepsilon$ as well as the lattice constants $a$ and $b$ by the double-Lorentzian fit on the intensity profile [Figs.~\ref{Fig4}(c) and \ref{Fig4}(d)].
$a$ and $b$ remains almost constant up to $\mu_{0}H_{\rm c1} \approx 0.75$~T, followed by an abrupt increase in $a$ and decrease in $b$.
On entering phase~II, $\varepsilon$ decreases to $1.0 \times 10^{-3}$ and gets even smaller with increasing the magnetic field.
Above $H_{\rm c2}$, the 400 peak splitting is invisible within instrument resolution, so that we plot the averaged lattice constant obtained from the single-Lorentzian fit on the intensity profile in Fig.~\ref{Fig4}(c).
The FWHM of the fitting function above $H_{\rm c2}$ represents a gradual decrease from phase~III to the FM phase [Fig.~\ref{Fig4}(d)], suggesting that slight orthorhombic distortion remains in phases~III and IV.
The correspondence between the crystal structure and spin textures with four-fold symmetry breaking is depicted in Fig.~\ref{Fig4}(f).

\begin{figure}[t]
\centering
\includegraphics[width=\linewidth]{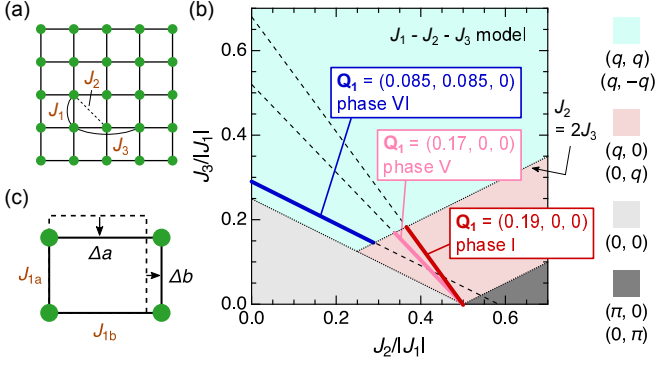}
\caption{(a) Exchange interactions up to the third NN on a square lattice. (b) Exchange parameter sets of $J_{2}$ and $J_{3}$ consistent with the fundamental ${\mathbf Q}$ vectors for phases VI, V, and I in zero field. The phase diagram is an excerpt from Ref.~\cite{2021_Wan} (see text for details). (c) Tetragonal-to-orthorhombic distortion on a square lattice, resulting in inquivalent NN exchanges $J_{1a} \approx J_{1}+(dJ_{1}/dr)\Delta a$ and $J_{1b} \approx J_{1}+(dJ_{1}/dr)\Delta b$, where $dJ_{1}/dr > 0$.}
\label{Fig5}
\end{figure}

{\it Discussions.} To understand the magnetic instability in EuAl$_{4}$, we start from a classical Heisenberg model on a simple square lattice: ${\mathcal{H}}=\sum_{{\langle i, j\rangle}_{k}} J_{k} {\mathbf S}_{i} \cdot {\mathbf S}_{j}+{\mathcal{H}_{\rm ex}}$.
Here, $J_{1} < 0$, $J_{2} > 0$, and $J_{3} > 0$ are the exchange interactions up to the third nearest neighbor (NN) [Fig.~\ref{Fig5}(a)], which originates from the RKKY interaction in EuAl$_{4}$.
This model can host various kinds of single-${\mathbf Q}$ and multiple-${\mathbf Q}$ spin textures depending on the extra terms ${\mathcal{H}_{\rm ex}}$; e.g., the double-${\mathbf Q}$ ML is stabilized in the presence of the single-ion anisotropy and compass anisotropy \cite{2021_Wan}, which would be relevant to phase VI in EuAl$_{4}$ \cite{2022_Tak}.
If we neglect ${\mathcal{H}_{\rm ex}}$, incommensurate ordering vectors ${\mathbf Q}=(\pm q, \pm q, 0)$ with $q = \arccos (-\frac{J_{1}}{2J_{2}+4J_{3}})$ are selected for $2J_{3} > J_{2}$ whereas ${\mathbf Q}=(\pm q, 0, 0)$ or $(0, \pm q, 0)$ with $q = \arccos (-\frac{J_{1}+2J_{2}}{4J_{3}})$ for $J_{2} > 2J_{3}$ \cite{2021_Wan}.
The exchange parameter sets that agree with the ${\mathbf Q}$ vectors for the zero-field phases VI, V, and I \cite{2022_Tak, 2021_Kan} are plotted in a parameter space of $J_{2}/|J_{1}|$ and $J_{3}/|J_{1}|$ in Fig.~\ref{Fig5}(b).
As phases VI, V, and I should compete with each other within a small energy scale, $(J_{2}/|J_{1}|, J_{3}/|J_{1}|) \approx (0.3, 0.15)$ is a reasonable parameter position for EuAl$_{4}$.
We note that the ${\mathbf Q}$ switching at $T_{\rm N3}$ and $T_{\rm N4}$ cannot occur within the frozen $J_{1}$--$J_{2}$--$J_{3}$ model.
Here we propose the spin-lattice coupling as a driving force to modify $J_{1}$ through a magnetostructural transition.
This mechanism is reasonable to consider in EuAl$_{4}$ on the basis of the observed exceptionally large thermal expansion and magnetostriction associated with the magnetic transitions ($\sim$$10^{-3}$) compared to those in other SkL-hosting chiral magnets \cite{2016_Pet} and Eu/Gd-based itinerant magnets ($10^{-5}$$\sim$$10^{-4}$) \cite{2006_Pro, 2019_Tak, 2021_Spa}.
This collective phenomenon is known as the spin Jahn--Teller effect, where the magnetic frustration is relieved by favoring one of the competing exchange interactions through the lattice distortion \cite{2011_Lac}.
The $B_{1g}$-type distortion is selected below $T_{\rm N3}$ \cite{2019_Shi}, indicating that the FM $J_{1}$ along the $a$ and $b$ axes become inequivalent: $J_{1a} \approx J_{1}+(dJ_{1}/dr)\Delta a$ and $J_{1b} \approx J_{1}+(dJ_{1}/dr)\Delta b$, respectively [Fig.~\ref{Fig5}(c)].
As the ${\mathbf Q}$ vector is oriented along the $b$ axis at 5~K according to the RXS experiment, FM coupling is stronger for the shorter $a$ axis, so that $dJ_{1}/dr > 0$.
This picture is compatible with the ${\mathbf Q}$ switching from (0.085, 0.085, 0) to (0.17, 0, 0) at $T_{\rm N3}$ because the total exchange energy can be reduced for the latter after the modification of $J_{1}$.
The additional first-order transition from phase V to I at $T_{\rm N4}$, where a modulation-period changes from $q = 0.17$ to 0.19 while conserving its orientation, suggests the competition of these two spiral states in the presence of the spin-lattice coupling; phase I is eventually stabilized as a ground state by enhancing the orthorhombic structural distortion and consequently increasing $J_{2}/|J_{1b}|$.
We note that a uniaxial stress should facilitate the magnetostructural transitions at $T_{\rm N3}$ and $T_{\rm N4}$ because the system can save the elastic energy, which is indeed observed as mentioned above (Fig. S3 \cite{SM} in the Supplemental Material).

The importance of the spin-lattice coupling is also corroborated from the strong correlation between the in-field spin textures and orthorhombic structural distortion.
In the rhombic SkL phase (II), two fundamental ${\mathbf Q}$ vectors are tilted from ${\mathbf Q}_{1}=(q, q, 0)$ and ${\mathbf Q}_{2}=(q, -q, 0)$ ($q \sim 0.085$) by $\theta_{q} \sim 5^{\circ}$ \cite{2022_Tak}.
As can be seen from Fig.~\ref{Fig1}(c), a large structural distortion $\varepsilon \sim 1.0 \times 10^{-3}$ exists in phase II like in phase V, suggesting the importance of the spin-lattice coupling on stabilizing the rhombic-${\mathbf Q}$ modulation.
It is worth referring that in a cubic chiral magnet FeGe 0.3$\%$ uniaxial strain deforms the triangular SkL by 20$\%$ owing to the anisotropic modulation of the DM interaction \cite{2015_Shi}.
In the rhombic SkL in EuAl$_{4}$, comparable structural distortion and SkL deformation are spontaneously induced, i.e., without applying the mechanical force.
Even in the square SkL phase (III), a moderate structural distortion $\varepsilon \sim 3 \times 10^{-4}$ is found [Fig.~\ref{Fig1}(c)], indicating that $J_{1a}$ and $J_{1b}$ remain inequivalent.
Such a deviation from the tetragonal symmetry in phase III might be observed as a slight difference in $q_{a}$ and $q_{b}$ in our RXS data [Fig.~\ref{Fig4}(e)].
Interestingly, a reentrant symmetry breaking of the spin texture is seen in the high-field rhombic VL phase (IV), though the orthorhombic structural distortion seems monotonically suppressed toward higher fields [Figs.~\ref{Fig4}(d) and \ref{Fig4}(e)].
The reason why the VL is more prone to deformation (in terms of spin textures) than the SkL is elusive at this stage.
The investigation on the anisotropic elastic property by means of ultrasonic measurements \cite{2014_Nii, 2015_Pet} would deepen our understanding on the SRSB of the SkL and VL in EuAl$_{4}$.
In addition, a theoretical framework incorporating local phonon modes or inequivalent $J_{1a}$ and $J_{1b}$ on the square lattice would be a promising approach to reproduce versatile magnetic phases in EuAl$_{4}$.

{\it Conclusion.} In summary, we have comprehensively revealed the crystal-structure changes of EuAl$_{4}$ associated with the field-induced phase transitions to address the microscopic origin of the SRSB of SkL.
The amplitudes of the orthorhombic structural distortion are quantitatively estimated for each magnetic phase.
We also unveil the correlation between magnetic modulation and the underlying crystal-structure distortion.
The appearance of two types of SkL phases should originate from the magnetic frustration in momentum space coupled with the lattice degrees of freedom.
Furthermore, the orthorhombic structural distortion accompanies a pronounced magnetovolume effect.
EuAl$_{4}$ would be an ideal playground to explore the tunability of the SkL phases by pressure as well as uniaxial stress.

{\it Acknowledgments.} The authors appreciate S. Hayami and H. Yoshimochi for fruitful discussions.
The authors appreciate S. Shimomura for providing the data of the lattice constants of EuAl$_{4}$ in Ref.~\cite{2019_Shi}.
This work was financially supported by the JSPS KAKENHI Grants-In-Aid for Scientific Research (Grants No. 19H01835, No. JP19H05826, No. 20H00349, No. 20J10988, No. 21H04440, No. 21H04990, No. 21K13876, No. 21K18595, No. 22H04965, No. 22K14010), PRESTO (Grant No. JPMJPR20B4), and Asahi Glass Foundation.
The resonant x-ray scattering experiment at PF was performed under the approval of the Photon Factory Program Advisory Committee (Proposal No.~2022G551).
M.G. was a postdoctoral research fellow of the JSPS.


\begin{thebibliography}{99}
\bibitem{comment1} We used different EuAl$_{4}$ crystals in the FBG and RXS experiments. In Fig.~\ref{Fig4}(d), the demagnetization correction is performed for the FBG data so that the transition fields $H_{\rm c1} \sim H_{\rm c4}$ match those obtained in the RXS experiment.
\bibitem{comment2} We adopt the Lorentzian function for the fundamental Bragg peaks and the Gaussian function for the magnetic Bragg peaks. In phases II$\sim$IV, we performed the $(H, K, 0)$ mesh scans in the vicinity of the magnetic Bragg reflections. For these two-dimensional (2D) data, we adopt the 2D-Gaussian fitting to estimate the peak positions and FWHMs for $q_{a}$ and $q_{b}$.
\end{thebibliography}
\end{document}


\title{Supplemental Material for\\``Rhombic skyrmion lattice coupled with orthorhombic structural distortion in EuAl$_{4}$''}

\author{Masaki Gen}
\email{masaki.gen@riken.jp}
\affiliation{Department of Advanced Materials Science, University of Tokyo, Kashiwa 277-8561, Japan}
\affiliation{RIKEN Center for Emergent Matter Science (CEMS), Wako 351-0198, Japan}

\author{Rina Takagi}
\affiliation{RIKEN Center for Emergent Matter Science (CEMS), Wako 351-0198, Japan}
\affiliation{Institute of Engineering Innovation, University of Tokyo, Tokyo 113-0032, Japan}
\affiliation{Department of Applied Physics, University of Tokyo, Tokyo 113-8656, Japan}
\affiliation{PRESTO, Japan Science and Technology Agency (JST), Kawaguchi 332-0012, Japan}

\author{Yoshito Watanabe}
\affiliation{Department of Advanced Materials Science, University of Tokyo, Kashiwa 277-8561, Japan}

\author{Shunsuke Kitou}
\affiliation{RIKEN Center for Emergent Matter Science (CEMS), Wako 351-0198, Japan}

\author{Hajime Sagayama}
\affiliation{Institute of Materials Structure Science, High Energy Accelerator Research Organization, Tsukuba 305-0801, Japan}

\author{Naofumi Matsuyama}
\affiliation{Institute for Solid State Physics, University of Tokyo, Kashiwa 277-8581, Japan}

\author{Yoshimitsu Kohama}
\affiliation{Institute for Solid State Physics, University of Tokyo, Kashiwa 277-8581, Japan}

\author{Akihiko Ikeda}
\affiliation{Department of Engineering Science, University of Electro-Communications, Chofu, Tokyo 182-8585, Japan}

\author{Yoshichika \={O}nuki}
\affiliation{RIKEN Center for Emergent Matter Science (CEMS), Wako 351-0198, Japan}

\author{Takashi Kurumaji}
\affiliation{Department of Advanced Materials Science, University of Tokyo, Kashiwa 277-8561, Japan}

\author{Taka-hisa Arima}
\affiliation{Department of Advanced Materials Science, University of Tokyo, Kashiwa 277-8561, Japan}
\affiliation{RIKEN Center for Emergent Matter Science (CEMS), Wako 351-0198, Japan}

\author{Shinichiro Seki}
\affiliation{RIKEN Center for Emergent Matter Science (CEMS), Wako 351-0198, Japan}
\affiliation{Institute of Engineering Innovation, University of Tokyo, Tokyo 113-0032, Japan}
\affiliation{Department of Applied Physics, University of Tokyo, Tokyo 113-8656, Japan}
\begin{abstract}

\end{abstract}

\maketitle

\onecolumngrid

\vspace{+1.0cm}
This Supplemental Material includes contents as listed below:\\
\\
Sec.~I.\hspace{4.0mm}Experimental details of the FBG technique\\
Sec.~II.\hspace{2.9mm}Raw data of magnetostriction curves measured along [100] for settings 1 and 2\\
Sec.~III.\hspace{1.5mm}Additional experimental data for settings~1 and 2\\
Sec.~IV.\hspace{1.9mm}FBG experiments on a nearly-detwinned crystal (setting~3)\\
Sec.~V.\hspace{3.1mm}Possible slight orthorhombic structural distortion above $T_{\rm N3}$\\

\vspace{+1.0cm}
\section{\label{Sec1}Experimental details of the FBG technique}

The dilatometry measurements were performed by the fiber-Bragg-grating (FBG) method using an optical sensing instrument (Hyperion si155, LUNA), which monitors the Bragg-wavelength changes ($\Delta \lambda/\lambda_{0}$) of the attached FBGs.
For conversion from the experimental $\Delta \lambda/\lambda_{0}$ to the sample-length change ($\Delta L/L_{0}$), we use the linear relation $\Delta \lambda/\lambda_{0} = c(\Delta L/L_{0})$ with a typical coupling parameter $c=0.76$ \cite{2010_Dao}, which should be satisfied well below 50~K where the thermal expansion of background components are negligible.
We confirm the validity of the above analysis in the previous work \cite{2022_Miy} and also in this study (see the main text).
Our FBG technique is capable of high-precision detection of $\Delta L/L_{0}$ with a resolution of $\sim$10$^{-7}$, and allows simultaneous multi-axis measurements \cite{2022_Miy, 2022_Gen}.

\vspace{+1.0cm}
\section{\label{Sec2}Raw data of magnetostriction curves measured along [100] for settings 1 and 2}

\begin{figure}[t]
\centering
\renewcommand{\thefigure}{S\arabic{figure}}
\vspace{+0.5cm}
\includegraphics[width=0.95\linewidth]{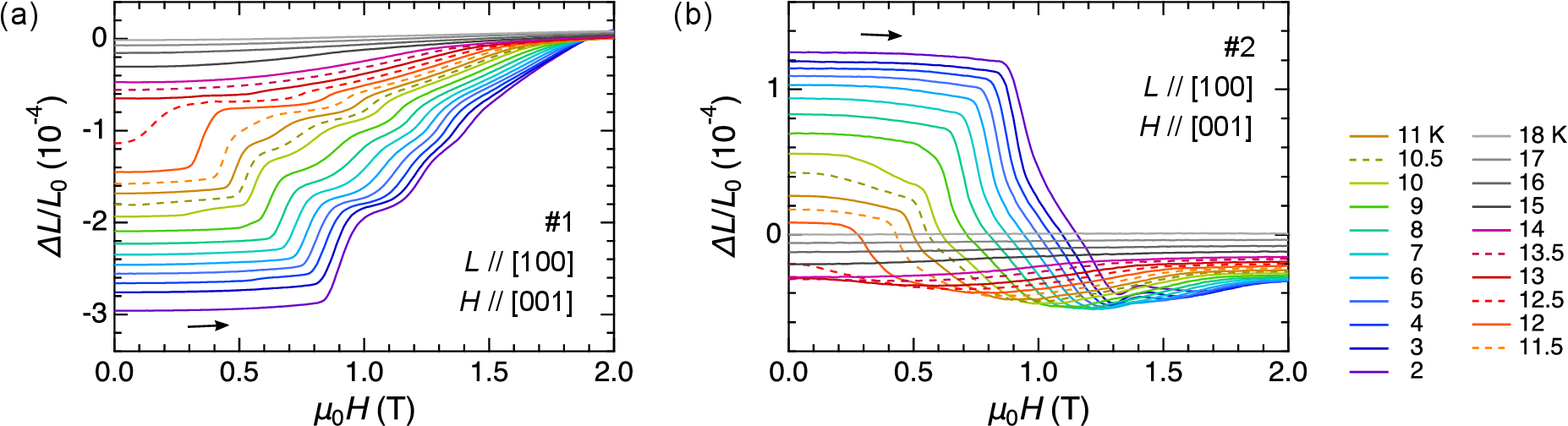}
\caption{Raw data of the observed magnetostriction curves for $H \parallel [001]$ and $\Delta L/L_{0} \parallel [100]$ in EuAl$_{4}$: (a) $(\Delta L/L_{0})_{1}$ for setting~1 and (b) $(\Delta L/L_{0})_{2}$ for setting~2. The color scale of temperature is shown in the right side of (b).}
\label{FigS1}
\vspace{+1.0cm}
\end{figure}

Figure~\ref{FigS1} shows the raw data of magnetostriction curves for $H \parallel [001]$ measured along [100] at various temperatures: (a) $(\Delta L/L_{0})_{1}$ for setting~1 and (b) $(\Delta L/L_{0})_{2}$ for setting~2.
Note that $(\Delta L/L_{0})_{1}$ and $(\Delta L/L_{0})_{2}$ were taken in independent sample settings, where the surfaces of the crystal and fibers were simply covered by Stycast1266 in a configuration shown in Fig.~2(a) in the main text.
By using the relations $\Delta a/a_{0} = [p_{2}(\Delta L/L_{0})_{1}-p_{1}(\Delta L/L_{0})_{2}]/(p_{2}-p_{1})$ and $\Delta b/b_{0} = [-(1-p_{2})(\Delta L/L_{0})_{1}+(1-p_{1})(\Delta L/L_{0})_{2}]/(p_{2}-p_{1})$ with $p_{1} = 0.49$ and $p_{2} = 0.74$, the field evolutions of $\Delta a/a_{0}$ and $\Delta b/b_{0}$ at the corresponding temperatures are obtained as shown in Figs.~3(a) and 3(b) in the main text.

\section{\label{Sec3}Additional experimental data for settings~1 and 2}

\begin{figure}[h]
\centering
\renewcommand{\thefigure}{S\arabic{figure}}
\vspace{+0.5cm}
\includegraphics[width=0.95\linewidth]{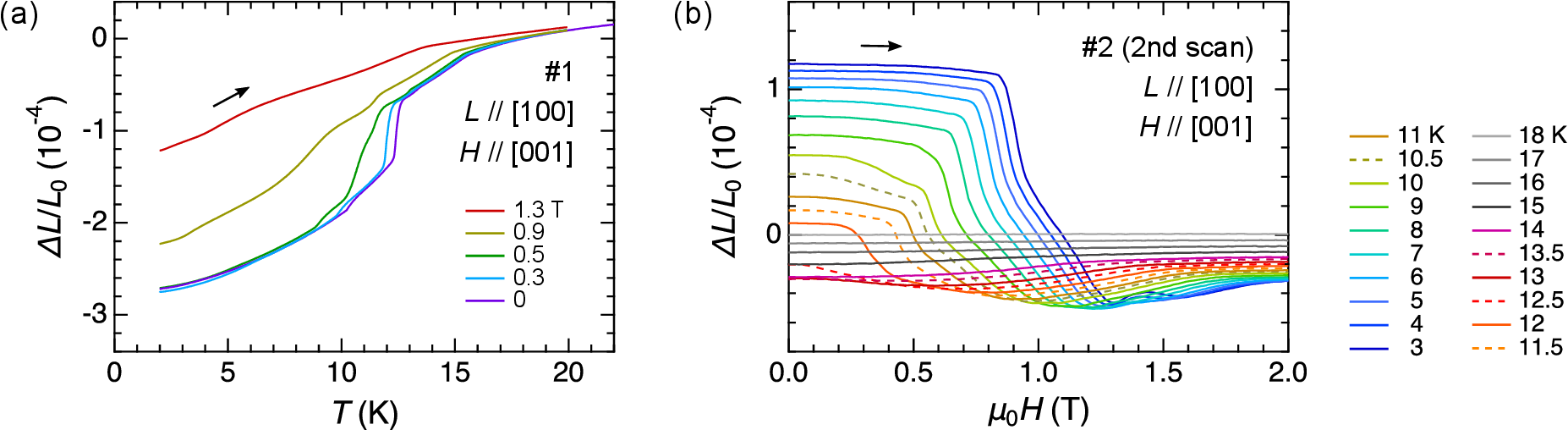}
\caption{(a) Thermal expansion measured along [100] at selected magnetic fields for $H \parallel [001]$ in the warming process for setting~1. (b) Magnetostriction measured along [100] at selected temperatures for $H \parallel [001]$ in the field-increasing process for setting~2.}
\label{FigS2}
\end{figure}

Figure~\ref{FigS2}(a) shows thermal expansion curves measured along [100] at constant magnetic fields for setting~1.
These data were obtained after a series of magnetostriction measurements shown in Fig.~\ref{FigS1}(a).
By comparing Figs.~\ref{FigS1}(a) and \ref{FigS2}(a), it can be seen that $\Delta L/L_{0}$ has the same value under certain temperature and magnetic field regardless of the sweep process.
This ensures that the volume ratio of the crystallographic domains never changes with scanning temperature as well as magnetic field.
Figure~\ref{FigS2}(b) shows magnetostriction curves measured along [100] at various temperatures for setting~2.
These data were obtained on a different day (i.e., the probe was once taken out from the superconducting magnet and inserted again) without changing the sample setting in which the data shown in Fig.~\ref{FigS2}(b) were obtained, but they were surprisingly reproducible.

Let us here propose the possible reasons for the above experimental observations. 
The energy barrier required to be overcome for the domain creation/annihilation as well as the domain-wall shift is expected to be significant in EuAl$_{4}$ because of the relatively large orthorhombic distortion ($\varepsilon \sim 10^{-3}$), the topological nature of spin textures, and the strong spin-lattice coupling.
Therefore, it would be natural to believe that no domain switching occurs as long as the crystal symmetry remains orthorhombic through the field-induced phase transitions.
We also expect that the adhesion of fibers to the sample surface by Stycast1266 should produce a certain amount of mechanical stress and that the way of domain distribution tends to be memorized.

\newpage
\section{\label{Sec4}FBG experiments on a nearly-detwinned crystal (setting~3)}

\begin{figure}[h]
\centering
\renewcommand{\thefigure}{S\arabic{figure}}
\vspace{+0.5cm}
\includegraphics[width=0.95\linewidth]{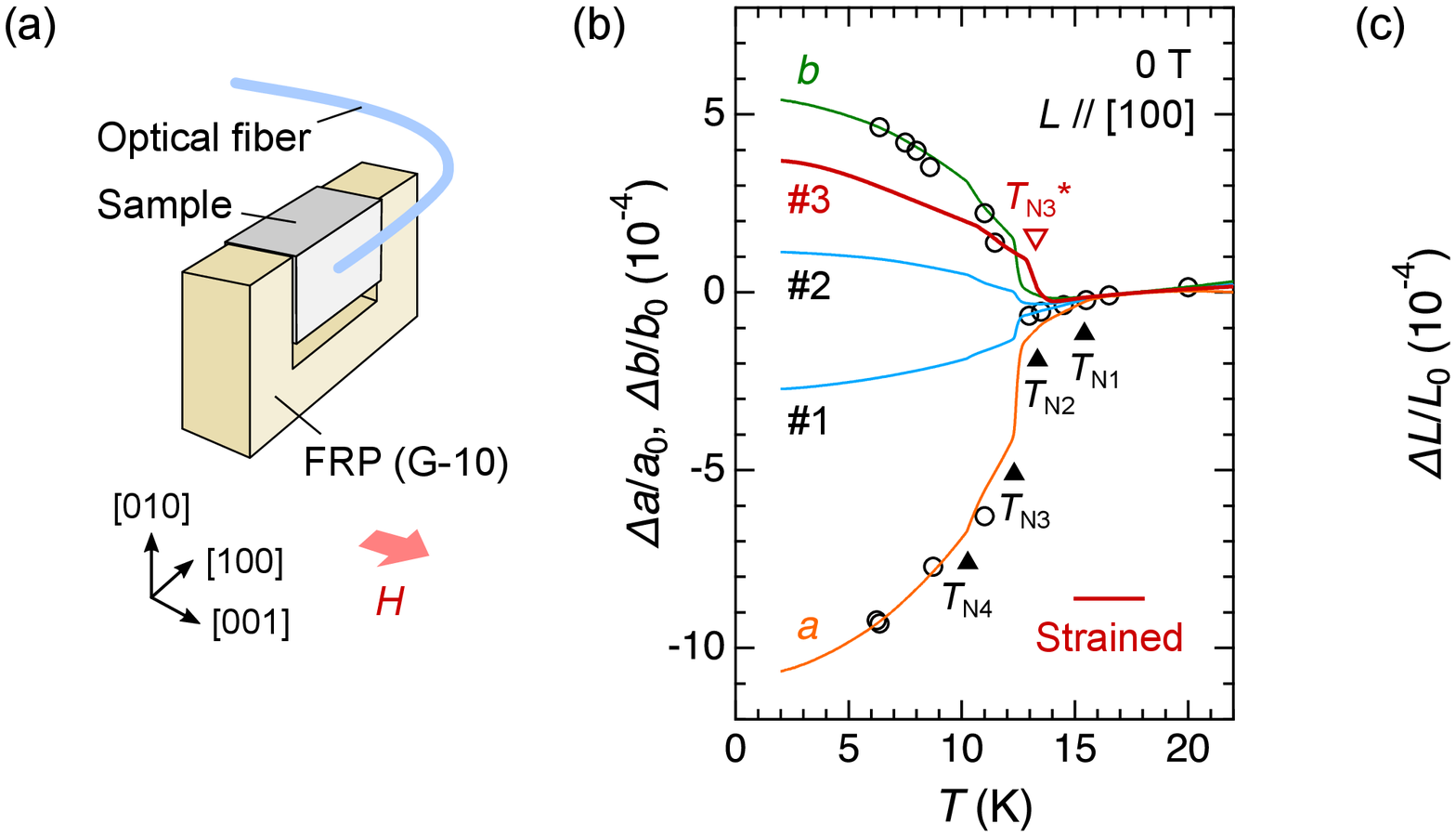}
\caption{(a) Schematic of the sample setting of the FBG experiments on a strained EuAl$_{4}$ crystal. The tensile-strain should be applied along the measured direction at low temperatures. (b) Thermal expansion measured along [100] in zero field on unstrained (data~\#1 and \#2 drawn by cyan lines) and strained crystals (data \#3 drawn by red line). See also Fig.~2(b) in the main text. (c) Magnetostriction curves for $H \parallel [001]$ and $\Delta L/L_{0} \parallel [100]$ on a strained crystal.}
\label{FigS3}
\end{figure}

In order to avoid the occurrence of crystallographic domains and directly detect the lattice constant change $\Delta b/b_{0}$ below $T_{\rm N3}$, we performed the FBG experiments by applying the thermal stress on a EuAl$_{4}$ single crystal.
The experimental setting is illustrated in Fig.~\ref{FigS3}(a).
The crystal was placed in a groove of the U-shaped block made of FRP (G-10), and facing (100) planes of the crystal were glued to the sides of the groove using Stycast1266.
The optical fiber was adhered to the (001) plane along the bridging [100] direction.
Since the thermal expansion coefficient of FRP is smaller than that of EuAl$_{4}$, the crystal should be tensile-strained along the bridging direction on cooling and could enter a nearly detwinned state below $T_{\rm N3}$.
Note that the similar way to apply the thermal stress has been previously employed in Ref.~\cite{2015_Shi}.

The observed thermal expansion profile in zero field is shown by a red line in Fig.~\ref{FigS3}(b).
A large negative thermal expansion is observed below $T_{\rm N3}^{*} \approx 13.5$~K, reaching $\Delta L/L_{0} \approx 4 \times 10^{-4}$ at 2~K.
By comparing with the previous x-ray diffraction (XRD) data indicated by open circles \cite{2019_Shi}, it is found that $\sim$90\% of $b$ component contributes to the elongation of the FBG.
Importantly, the original magnetostructural transition temperature $T_{\rm N3} = 12.2$~K shifts higher by more than 1~K, suggesting that the magnetism in EuAl$_{4}$, which should be strongly coupled to the lattice degrees of freedom, is sensitive to the uniaxial stress.
Figure~\ref{FigS3}(c) shows the magnetostriction curves for $H \parallel [001]$ measured along [100] at various temperatures on the strained crystal.
These data are reasonably consistent with the estimated $\Delta a/a_{0}$ and $\Delta b/b_{0}$ shown in Figs.~3(a) and 3(b) in the main text.
Note that we could not confirm the shift of the critical fields $H_{\rm c1} \sim H_{\rm c4}$ from the original values at 2~K.
It would be necessary to apply a stronger mechanical stress to control the stability of each SkL phase.

\vspace{+1.0cm}
\section{\label{Sec5}Possible slight orthorhombic structural distortion above {\textit T}$_{\bf N3}$}

As for the estimation of the orthorhombic structural distortion $\varepsilon \equiv (b - a)/a_{0}$ shown in Fig.~1(c), if we simply assume $a_{0} = b_{0}$ at 0~T and 18~K, $\varepsilon$ turns to negative above $\sim$1.7~T at 2~K.
This would not be intrinsic but presumably due to the missing incorporation of a tiny distortion at 0~T and 18~K, i.e., $a_{0} < b_{0}$, in our analysis.
The recent XRD study by Ramakrishnan {\it et al.} \cite{2022_Ram} suggested the presence of orthorhombic distortion as much as $\sim$$1 \times 10^{-4}$ within the $ab$ plane below $\sim$140~K, where the charge density wave (CDW) with an incommensurate modulation along the $c$ axis emerges \cite{2015_Nak, 2019_Shi}.
They argue that the in-plane lattice distortion is the $B_{2g}$ type ($Fmmm$ symmetry) rather than the $B_{1g}$ type ($Immm$ symmetry) \cite{2022_Ram}.
However, we observe that $\Delta b/b_{0}$ deviates from $\Delta a/a_{0}$ between $T_{\rm N3}$ and $T_{\rm N1}$ [Fig.~2(b)], indicating the presence of a slight $B_{1g}$-type lattice distortion above $T_{\rm N3}$.
We infer that the both the $B_{1g}$ and $B_{2g}$ modes may be relevant in the CDW state, which would affect the estimation of $\varepsilon$ especially in high-field or high-temperature sides.
This ambiguity does not affect the conclusion of the present work.